\documentclass[twocolumn,aps,prl,superscriptaddress,preprintnumbers,amsmath,amssymb,nofootinbib]{revtex4-2}
\usepackage{epsfig}  
\usepackage{graphicx}
\usepackage{color}
\usepackage[dvipsnames]{xcolor}
\usepackage{float}
\usepackage{placeins}
\usepackage{tabularx} 
\usepackage{amsfonts}
\usepackage{amsmath}
\usepackage{mathrsfs}
\usepackage{slashed}
\usepackage{soul}
\usepackage{braket}
\usepackage{dsfont}
\usepackage[colorlinks,citecolor=blue,urlcolor=magenta,bookmarks=false,hypertexnames=true]{hyperref}
\usepackage{orcidlink}
\usepackage{multirow}
\usepackage{colortbl}
\usepackage{graphicx}
\usepackage{dcolumn}

\large

\begin{document}

\title{Experimental limits on quantum decoherence from $B$ meson systems}

\author{Ashutosh Kumar Alok}
\thanks{\textcolor{magenta}{Deceased}}
\affiliation{Indian Institute of Technology Jodhpur, Jodhpur 342037, India}

\author{Neetu Raj Singh Chundawat}
\email{chundawat.1@alumni.iitj.ac.in}
\affiliation{Indian Institute of Technology Jodhpur, Jodhpur 342037, India}
\affiliation{Institute of High Energy Physics, Chinese Academy of Sciences, Beijing 100049, China}

\author{Jitendra Kumar}
\email{jkumar@iitj.ac.in}
\affiliation{Indian Institute of Technology Jodhpur, Jodhpur 342037, India}

\author{Saurabh Rai}
\email{saurabhrai@iitj.ac.in}
\affiliation{Indian Institute of Technology Jodhpur, Jodhpur 342037, India}

\author{S. Uma Sankar}
\email{uma@phy.iitb.ac.in}
\affiliation{Indian  Institute  of  Technology  Bombay,  Mumbai  - 400076, India}

\begin{abstract}
Neutral $B$-meson systems serve as critical tests of the Standard Model and play a key role in limiting its extensions. While these systems are typically studied under the assumption of perfect quantum coherence, interactions with the environment can lead to decoherence. Such decoherence effects  
can obscure the measured values of key parameters such as the oscillation frequency \( \Delta m \) and $CP$-violating parameter \( \sin 2\beta \). Using the experimental data, we present the \textit{first} combined analysis of mixing asymmetry and $CP$-asymmetry measurements for \( B_d \)-mesons, which indicates that \( \lambda_d \) is non-zero at approximately \( 6 \,\sigma \). We also establish the \textit{first} experimental constraints on the decoherence parameter \( \lambda_s \) for \( B_s \)-mesons, finding it to be non-zero at \( 3 \,\sigma \).

\end{abstract}

\maketitle
\newpage

\textit{Introduction.}---Flavor physics offers an innovative approach to exploring physics beyond the Standard Model (SM) by using precise measurements to detect the virtual production of new particles in quantum loops, probing energy scales far beyond the reach of direct high-energy physics experiments. In particular, the study of $B$-meson decays has subjected the SM of electroweak interactions to stringent tests and placed strong constraints on numerous extensions beyond the SM. With the lack of direct evidence for new physics searches in high-energy physics experiments like ATLAS and CMS, the importance of flavor physics experiments, such as LHCb and Belle II, has grown even more critical.

The time evolution of neutral mesons determines several essential parameters in flavor physics. It is generally assumed that the quantum coherence inherent in the system remains perfect during this evolution. However, in practice, all systems interact with their environment, and such interactions can lead to a loss of quantum coherence. 
Decoherence is inevitable, as any physical system is inherently open due to its unavoidable interactions with the surrounding environment. These environmental effects can originate from fundamental sources, such as fluctuations in the quantum gravity space-time background \cite{Hawking:1979pi,noncritical, Mavromatos:2004sz, Gambini:2003pv}. The impact of the environment on neutral-meson systems can be effectively incorporated using the framework of open quantum systems \cite{Lindblad:1975ef,Gorini:1975nb,Gorini:1976cm,Matthews:1961vr,kraus,alicki, breuer,Caban:2007je,Caban:2005ue}. This formalism allows for modeling the time evolution of neutral mesons while inherently accounting for decoherence effects. Such an approach offers an effective phenomenological description that does not rely on the specific details of the interactions between the system and its environment.

In the open quantum systems framework, neutral $B$-mesons are treated as subsystems interacting with an environment. The standard unitary operator governs the entire system's evolution, while the $B$-mesons' dynamics alone are derived by integrating out the environmental degrees of freedom. Assuming the interaction between the $B$-meson and the environment is weak, the system's dynamics can be described using quantum dynamical semigroups, which satisfy the condition of complete positivity. We use the density matrix formalism to describe the time evolution of the neutral $B$-meson system \cite{Blasone:2007wp,Banerjee:2014vga,Naikoo:2018vug,Dixit:2018gjc,Alok:2015iua}. The time evolution of these matrices can then be expressed in an operator-sum form with the help of Kraus operators \cite{Caban:2007je}.

The Kraus operators, initially introduced for the $K$ meson system, have been utilized in \cite{Alok:2015iua} to explore the $B_q (q = d,s)$ meson systems. It was shown that the inclusion of decoherence effects can mask the measured values of some key parameters such as oscillation frequency $\Delta m_q$, \(CP\)-violation parameter $\sin 2 \beta_q$, and decay width difference $\Delta \Gamma_q$. Therefore, the data used in the measurement of these observables can also be used to determine bounds on the decoherence parameter.
In Ref.~\cite{Alok:2024amd}, authors demonstrate the analysis of several observables in semileptonic and non-leptonic decays of \(B\)-mesons, which can also be utilized to determine the decoherence parameter. 

In this work, we use published data from the experiments and demonstrate how the combined analysis of time-dependent mixing asymmetry and $CP$ asymmetry, used to determine the $B^0_d - \bar{B}^0_{d}$ oscillation frequency $\Delta m_d$ and other $CP$-violating parameters can be employed to establish bounds on the decoherence parameter. We also obtain the first experimental limits on the decoherence parameter derived from the $B_s$-meson system using the mixing asymmetry measurements. 


\textit{Mixing asymmetry in $B^0_q$-meson system.}---A neutral \( B^0_q \)-meson can oscillate into a \( \bar{B}^0_q \)-meson, denoted as \( B^0_q \leftrightarrow \bar{B}^0_q \). 
The Flavor-Changing Neutral Current (FCNC) can only occur at the loop level within the SM due to the Glashow-Iliopoulos-Maiani (GIM) mechanism. Consequently, \( B^0_q - \bar{B}^0_q \) oscillations in the SM happen exclusively through loop-level {\it box} diagrams, making the process sensitive to the presence of heavy particles beyond the SM. The first observation of \( B^0_q -\bar{B}^0_q \) mixing by the ARGUS collaboration in 1987 provided indirect evidence for a heavy top quark, which was eventually discovered in 1995 by the CDF and D$\O$  collaborations. 

In the presence of decoherence, the probability for an initial \( B^0_q \)-meson to survive as \( B^0_q \) or oscillate into \( \bar{B}^0_q \) at a given proper decay time \( t \) is expressed as follows:
\begin{align}
P_{+}(t, \lambda_q) &= \frac{e^{-\Gamma t}}{2} \left[ \cosh\left( \frac{\Delta \Gamma_q t}{2} \right) + e^{-\lambda_q t} \cos(\Delta m_q \, t) \right], \nonumber \\
P_{-}(t, \lambda_q) &= \frac{e^{-\Gamma t}}{2} \left| \frac{\mathbf{q}}{\mathbf{p}} \right|^2 \left[ \cosh\left( \frac{\Delta \Gamma_q t}{2} \right) - e^{-\lambda_q t} \cos(\Delta m_q \, t) \right],
\label{mixprob}
\end{align}
where \( \Gamma = (\Gamma_L + \Gamma_H)/2 \) represents the average decay width, and \( \Delta \Gamma_q = \Gamma_L - \Gamma_H \). \( \Gamma_L \) and \( \Gamma_H \) are the decay widths of \( B^0_L = \mathbf{p} B^0 + \mathbf{q} \bar{B}^0 \) and \( B^0_H = \mathbf{p} B^0 - \mathbf{q} \bar{B}^0 \), respectively, which correspond to the light and heavy states of the \( B \)-mesons. The complex coefficients \( \mathbf{p}\) and \( \mathbf{q} \) satisfy the condition \( |\mathbf{p}|^2 + |\mathbf{q}|^2 = 1 \). Here, $\lambda_q$ is the decoherence parameter. Also, \( P_{+}(t, \lambda_q) \) corresponds to the case where the meson decays as the same flavor as its initial state, while \( P_{-}(t, \lambda_q) \) applies when it decays as the opposite flavor. In the limit of neglecting the $CP$-violation in mixing, the expression for the time-dependent mixing asymmetry \( A_{\rm mix}(t, \lambda_q) \), defined as \( A_{\rm mix} (t, \lambda_q) \equiv \frac{P_{+} (t, \lambda_q) - P_{-} (t, \lambda_q)}{P_{+} (t, \lambda_q) + P_{-} (t, \lambda_q)} \), through which the $B^0_q$ oscillation frequency $\Delta m_q$ is determined, is 
\begin{equation}
    A_{\rm mix} (t,\lambda_q)=
\frac{ \cos (\Delta m_q\, t) }{\cosh (\Delta \Gamma_q\, t/2) } e^{-\lambda_q\, t}\,.
\label{amix}
\end{equation}
Thus, it is evident from the above equation that the mixing asymmetry is influenced by the decoherence, represented by the factor \( e^{-\lambda_q t} \).

Over the past two decades, numerous time-dependent \( B_d^0 \)–\( \bar{B}_d^0 \) oscillation analyses have been conducted by the ALEPH, DELPHI, L3, OPAL, CDF, D0, BABAR, Belle, Belle II, and LHCb collaborations. However, none of these analyses took the decoherence into account. In colliders-based experiments such as Tevatron and LHC, usually the decay rates of particles, initially prepared as pure \( B^0_d \) states at \( t = 0 \), which later decay as \( B^0_d \) or \( \bar{B}^0_d \) over their proper decay time \(t\), are measured. This proper decay time is determined using the distance between the production vertex and the neutral $B$ decay vertex along with the momentum of the reconstructed $B$-meson. In the case of asymmetric colliders-based experiments such as BaBar, Belle or Belle II, the time-dependent probabilities \( P_+(t, \lambda_{q}) \) for un-oscillated \( B^0_d \bar{B}^0_d \) events and \( P_-(t, \lambda_{q}) \) for oscillated \( B^0_d B^0_d \) or \( \bar{B}^0_d \bar{B}^0_d \) events are measured and the proper time difference \( \Delta t \) between the two neutral $B$-mesons is used. The current world average of \( B_d \)-mesons oscillation frequency $\Delta m_d$ under the assumptions of \( \Delta \Gamma_d = 0 \) and no $CP$ violation in mixing yield \( \Delta m_d = 0.5069\,\pm\,0.0019 \, \mathrm{ps}^{-1} \) \cite{HFLAV:2022esi}.

The measurement of the proper decay time and a precise decay time resolution is even more crucial in $B_s$ meson system as $B_s^0 - \bar{B_s^0}$ oscillations are approximately 35 times faster than the $B_d^0 - \bar{B_d^0}$ oscillations. The investigation of the neutral \( B_s \) system has become more feasible due to advancements in hadron colliders such as the LHC, following earlier studies at the Tevatron. The LHCb collaboration reported several measurements for \( B_s^0 - \bar{B}_s^0 \) oscillations using Run 1 and Run 2 data~\cite{LHCb:2011vae, LHCb:2013lrq, LHCb:2013fep, LHCb:2014iah, LHCb:2020qag, LHCb:2021moh, LHCb:2023sim}. CMS collaboration also reported their results based on the Run 2 dataset~\cite{CMS:2020efq}. Here again, the decoherence is not included while determining the mass difference. The current world average values of the oscillation frequency and decay width difference in \( B_s \)-mesons are given as, \( \Delta m_s = 17.765 \pm 0.004 (stat.) \pm 0.004 (syst.) \, \mathrm{ps^{-1}} \)~\cite{HFLAV:2022esi} and $\Delta \Gamma_s = 0.083 \pm 0.005 \, \mathrm{ps^{-1}}$~\cite{ParticleDataGroup:2024cfk}, respectively.

As we see from Eq.~\eqref{amix}, the determination of \( \Delta m_q \) and \(  \Delta\Gamma_q \) is influenced by the decoherence effects. Therefore, re-evaluating the time-dependent mixing asymmetry data for \( B_q^0 - \bar{B}_q^0 \) oscillations can help to extract the \textbf{true} values of  \( \Delta m_q \) and \( \Delta \Gamma_q \), while simultaneously providing a means to place constraints on the decoherence parameter \(\lambda_q\). A first attempt to estimate the decoherence parameter $\lambda_d$ was made using mixing asymmetry data in Ref.~\cite{Alok:2015iua}. 
\textit{\( CP\) asymmetry in $B^0_q$-meson system.}---Apart from the oscillation effects discussed so far, neutral meson systems also exhibit the phenomenon of \( CP \)-violation. These mesons can decay into states of definite $CP$ (denoted as \( f_{\text{CP}} \)).  We define the amplitude for the decay \( B^0_q \to f_{\text{CP}} \) as \( A_f \), and the amplitude for the decay \( \bar{B}^0_q \to f_{\text{CP}} \) as \( \bar{A}_f \). It is standard to introduce the parameter:
\[ \lambda_{f_{CP}} = \frac{\bar{A}_f}{A_f} \frac{\mathbf{q}}{\mathbf{p}},\]
where \( \mathbf{q} \) and \( \mathbf{p} \) are the complex coefficients related to \( B^0_q - \bar{B}^0_q \) mixing. The probability rate for an initial state \( B^0_q \) or \( \bar{B}^0_q \) decaying into a final state \( f_{CP} \) is given by:
\begin{eqnarray}
    \frac{P_{f_{CP}}(B^0_q;t,\lambda_q)}{\frac{1}{2} e^{- \Gamma t}|A_f|^2} &=&
\left(1+|\lambda_{f_{CP}}|^2\right) \cosh \left(\frac{\Delta \Gamma_q t}{2}\right) \nonumber\\
&+&\left(1-|\lambda_{f_{CP}}|^2\right)e^{- \lambda_q t} \cos\left(\Delta m_q t\right) \nonumber\\ 
&-& 2\,{\rm Re}(\lambda_{f_{CP}})\sinh \left(\frac{\Delta \Gamma_q t}{2}\right) \nonumber\\
&-& 2\,{\rm Im}(\lambda_{f_{CP}})e^{- \lambda_q t} \sin\left(\Delta m_q t\right),
\end{eqnarray}
\begin{eqnarray}
\frac{P_{f_{CP}}(\bar{B}^0_q;t,\lambda_q)}{\frac{1}{2} e^{- \Gamma t}|A_f|^2 \left|\frac{\mathbf{p}}{\mathbf{q}}\right|^2} &=& 
\left(1+|\lambda_{f_{CP}}|^2\right) \cosh \left(\frac{\Delta \Gamma_q t}{2}\right)\nonumber\\
&-&\left(1-|\lambda_{f_{CP}}|^2\right)e^{- \lambda_q t} \cos\left(\Delta m_q t\right) \nonumber \\ 
&-& 2\,{\rm Re}(\lambda_{f_{CP}})\sinh \left(\frac{\Delta \Gamma_q t}{2}\right)\nonumber\\
&+& 2\,{\rm Im}(\lambda_{f_{CP}})e^{- \lambda_q t} \sin\left(\Delta m_q t\right).
\end{eqnarray}

From these expressions, the $CP$-violating observable is defined as:
\begin{equation}
\mathcal{A}_{f_{CP}}(t,\lambda_q) = \frac{P_{f_{CP}}(B^0_q;t,\lambda_q) - P_{f_{CP}}(\bar{B}^0_q;t,\lambda_q)}{P_{f_{CP}}(B^0_q;t,\lambda_q) + P_{f_{CP}}(\bar{B}^0_q;t,\lambda_q)}.
\label{acpexp}
\end{equation}

Neglecting $CP$ violation in mixing, the expression of time-dependent $CP$ asymmetry in the presence of decoherence is given by
{\small \begin{equation}
{\mathcal{A}}_{f_{CP}} (t,\lambda_q) = 
\frac{A_{\rm CP}^{\rm dir,\,f_{CP}}  \cos\left(\Delta m_q t\right)+ A_{\rm CP}^{{\rm mix},\,f_{CP}} \sin\left(\Delta m_q t\right)}
{ \cosh \left(\frac{ \Delta \Gamma_q t}{2}\right)+A_{\Delta \Gamma}^{f_{CP}}  \sinh \left(\frac{ \Delta \Gamma _qt}{2}\right)} e^{-  \lambda_q t} \,,
\label{cpasym1}
\end{equation}}
where
\begin{eqnarray}
    A_{\rm CP}^{\rm dir,\,f_{CP}} &=& \frac{1-|\lambda_{f_{CP}}|^2}{1+|\lambda_{f_{CP}}|^2},  \nonumber\\ A_{\Delta \Gamma}^{f_{CP}} &=& - \frac{2 {\rm Re}(\lambda_{f_{CP}})}{1+|\lambda_{f_{CP}}|^2}, \nonumber\\  A_{\rm CP}^{{\rm mix},\,f_{CP}} &=& -\frac{2 {\rm Im}(\lambda_{f_{CP}})}{1+|\lambda_{f_{CP}}|^2}\,.
\end{eqnarray}

It is important to note that the above observables are related by  
\( (A_{\rm CP}^{\rm dir,\,f_{CP}})^2 + (A_{\rm CP}^{{\rm mix},\,f_{CP}})^2 + (A_{\Delta \Gamma}^{f_{CP}})^2 = 1 
\) relation and are masked by the presence of decoherence parameter \(\lambda_q\) along with $\Delta m_q$ and $\Delta \Gamma_q$. The time-dependent \( CP \) asymmetry, as shown in Eq.~\eqref{cpasym1} with \(\lambda_q = 0\), is a crucial tool for extracting parameters like \( \sin 2\beta \) and \( \sin 2\beta_s \). The angles \( \beta \equiv \arg\left[ -(V_{cd} V_{cb}^*) / (V_{td} V_{tb}^*) \right] \) and \( \beta_s \equiv \arg\left[ -(V_{ts} V_{tb}^*) / (V_{cs} V_{cb}^*) \right] \) are key elements of the unitarity triangle.

A precise determination of \( \sin 2\beta \) is achieved through a time-dependent angular analysis of the decay \( B^0_d \to J/\psi\, K_S \), commonly regarded as the ``golden channel" for this measurement. A final state that serves as a \(CP\) eigenstate is necessary, one to which both \( B^0_d \) and \( \bar{B}^0_d \) can decay. This decay mode has the quark-level transition \( b \to c\bar{c}s \), which enables access to the angle \( \beta \). For this mode, where \( \Delta \Gamma_d = 0 \), \( \lambda_{f_{CP}} = 1 \), and the penguin loop contributions to \( b \to c\bar{c}s \) are negligible, the \( CP \)-asymmetry simplifies to \( \mathcal{A}_{f_{CP}}(t) \approx  e^{-\lambda_d t} \sin 2\beta \cos\left(\Delta m_d t\right)\). This demonstrates that the coefficient of \( \sin(\Delta m_d t) \) in the \( CP \) asymmetry is \( e^{-\lambda_d t} \sin 2\beta \), not simply \( \sin 2\beta \), highlighting the modulation introduced by the decoherence factor \( e^{-\lambda_d t} \). 

\textit{Estimation of decoherence parameter from \( B_d\) system.}--- The time-dependent mixing asymmetry and $CP$ asymmetry measurements in the \( B_d \)-meson system are affected by the decoherence effect. Therefore, conducting a combined fit to the available data would provide the most robust and reliable constraint. We conducted a comprehensive review of all reported experimental results to date. We have considered mixing asymmetry results where data is available more precisely for larger decay times as it is essential to study the decoherence parameter. As a result, most of the data from the BaBar~\cite{BaBar:2001qlo, BaBar:2002epc, BaBar:2001bcs, BaBar:2002jxa, BaBar:2005laz} and Belle~\cite{Belle:2002qxn, Belle:2002lms, Belle:2004hwe} collaborations cannot be fully utilized. The LHCb data~\cite{LHCb:2016gsk} derived from measurements of the oscillation frequency of \( B_d \)-mesons in the decays \( B_d^0 \to D^{(*)-}\mu^+\nu_\mu X \) and \( B_d^0 \to D^-\mu^+\nu_\mu X \) is used in our analysis. In Ref.~\cite{LHCb:2016gsk}, the mixing asymmetry results were reported separately for the 2011 and 2012 runs, further categorized into four different tagging quality levels for each decay mode, resulting in a total of sixteen mixing asymmetry results. In our analysis, data with the best tagging quality from both decay modes with 2011 and 2012 runs are used, resulting in a total of four datasets. For \( CP \) asymmetry measurements, we use the latest LHCb measurement~\cite{LHCb:2023zcp} based on \( B^0_d \to \psi K_S^0 \) decays. It is to be noted that the LHCb uses an inverted convention here as compared to our definition in Eq.~\eqref{acpexp}.  The \texttt{WebPlotDigitizer}~\cite{WebPlotDigitizer} is utilized to extract numerical values from experimental results. 

\begin{table*}[htb]
\centering
\begin{tabular}{|c|c|c|c|c|} \hline \hline
Parameters & Values from LHCb& Best fit values with $\lambda_{d}=0$   & Best fit values with $\lambda_{d} \neq 0$   \\ 
 
 & & ($\chi^{2}_{bf}/dof =2.84$) & ($\chi^{2}_{bf}/dof=1.76$) \\
\hline
\hline
{$\Delta m_d$ (ps$^{-1}$)} & $0.505 \pm 0.002 (stat.) \pm 0.001 (syst.)$ \cite{LHCb:2023zcp} & $0.494 \pm 0.007$  & $0.469 \pm 0.005$    \\ 
\hline
$A_{\rm CP}^{{\rm dir},\,f_{CP}}$ & $-0.004 \pm 0.012$ \cite{LHCb:2016gsk} & $-0.010 \pm 0.018$ & $-0.005 \pm 0.021$ \\ 
\hline
$A_{\rm CP}^{{\rm mix},\,f_{CP}}$ & $0.724 \pm 0.014$ \cite{LHCb:2016gsk} & $0.711 \pm 0.020$ & $0.836 \pm 0.038$   \\ 
\hline
$\lambda_{d}$ (ps$^{-1}$) & $ --$ & $--$ & \textit{$0.055 \pm 0.009$}  \\ 
\hline
\end{tabular}
\caption{Comparison of parameter values from LHCb measurements and the best-fit values obtained from our fit for $B_d$ meson system.}
\label{tab:fit_results}
\end{table*}

The decay width difference between the \( B_d \)-meson mass eigenstates, \( \Delta \Gamma_d \), is assumed to be zero, and thus, the expressions in Eq.~\eqref{amix} and~\eqref{cpasym1} for the time-dependent mixing asymmetry and \( CP \)-asymmetry simplifies to their numerators only. We perform a chi-square (\( \chi^2 \)) fit on the data to determine the best-fit values of the parameters \( A_{\rm CP}^{\rm dir,\,f_{CP}} \), \( A_{\rm CP}^{{\rm mix},\,f_{CP}} \), and \( \Delta m_d \), assuming a no-decoherence scenario. The results are then compared with published values to validate the reliability of our fitting procedure. The fit quality is then monitored through reduced chi-squared statistic (\(\chi^2_{bf}\)/$dof$). The same procedure is then repeated, incorporating the effects of decoherence.

Our findings are summarized in Table~\ref{tab:fit_results}. For the combined fit performed without considering decoherence, we observe that all parameters align well with the corresponding LHCb measurements. For example, in the published results, \( \Delta m_d \) ranges from \( 0.497 \pm 0.006 \) ps$^{-1}$ to \( 0.508 \pm 0.004 \) ps$^{-1}$, depending on the run period and decay mode. Combining all data, \( \Delta m_d \) is found to be \( 0.505 \pm 0.002 (stat.) \pm 0.001 (syst.) \) ps$^{-1}$. Despite using only partial data, we obtain a consistent value of \( \Delta m_d = 0.494 \, \pm 0.007 \, \) ps$^{-1}$ for the best-fit values, confirming the reliability of the extracted data. The direct \( CP \) asymmetry \( A_{\rm CP}^{\rm dir,\,f_{CP}} \) and the mixing asymmetry \( A_{\rm CP}^{{\rm mix},\,f_{CP}} \) exhibit larger uncertainties compared to \( \Delta m_d \), primarily due to limited statistics in the corresponding experimental data. 
It is important to emphasize that \( \Delta m_d \)  is determined exclusively from the mixing asymmetry measurements and the \( CP \)-violating parameters are estimated from the $CP$-asymmetry, while the oscillation frequency (\(\Delta m_d\)) is fixed to its world average. In our analysis, we have done a simultaneous determination of \( \Delta m_d, A_{\rm CP}^{\rm dir,\,f_{CP}} \) and \( A_{\rm CP}^{{\rm mix},\,f_{CP}} \) by doing a fit to both $A_{\text{mix}}(t,\lambda_d)$ and $A_{f_{CP}}(t,\lambda_d)$ data.

When decoherence is included in the analysis, we observe that the central value of \( \Delta m_d \) shifts to a lower value and that of \( A_{\rm CP}^{{\rm mix},\,f_{CP}} \) shifts to a higher value by nearly $4\,\sigma$,  while \( A_{\rm CP}^{\rm dir} \) is essentially unchanged. The best-fit value for the decoherence 
parameter $\lambda_d = 0.055 \pm 0.009$, is
non-zero at $6\,\sigma$! This scenario offers an improved fit to the data, with \( \chi^2_{\rm bf}/{dof} = 1.76 \), compared to 2.84 in the earlier case. 
When direct \( CP \)-asymmetry is neglected i.e., \( A_{CP}^{\text{dir}, \, f_{CP}} \) is set to zero in the fit, the resulting values of remaining parameters are similar to the \(\lambda_d \neq 0\) case. 
Our fit shows strong evidence for a non-zero value of $\lambda_d$, which also shifted the values of the important physical parameters significantly. 
To determine the decoherence parameter in experiments, it is essential to also examine additional asymmetries defined in Ref.~\cite{Alok:2024amd}, which can provide clearer insights into whether the decoherence effect is non-zero. Hence, efforts in this direction from the experimental side, utilizing exact and up-to-date data, hold immense potential to precisely determine the value of the decoherence parameter. 

\textit{Estimation of decoherence parameter from \( B_s\) system.}--- In this case, most of the mixing asymmetry measurements are presented as a function of \( t \, \text{modulo} \, (2\pi/\Delta m_s) \) rather than the \( B_s \) decay time \( t \), including the most recent result from the LHCb collaboration in 2022 \cite{LHCb:2021moh}. While the asymmetry could, in principle, be directly constructed using the proper time distributions of the decays, extracting the necessary data from Ref.~\cite{LHCb:2021moh} is challenging. Therefore, we exclude this measurement from our analysis. To showcase the ability of this system to estimate the decoherence parameter, we utilize the available mixing asymmetry data from \( B^0_s - \bar{B}^0_s \) oscillation frequency measurements provided by the LHCb collaboration in 2013~\cite{LHCb:2013fep}. This measurement was performed with $B_{s}^{0} \to D_{s}^{-}\mu^{+}X$ decays over integrated luminosity around 1.0 fb$^{-1}$. Although the $B_{s}^{0}$ oscillations were detected with a significance of $5.8 \,\sigma$, they were observed to gradually diminish, an effect likely due to poor decay time resolution.
\begin{figure}[htb]
\centering
\includegraphics[scale=0.43]{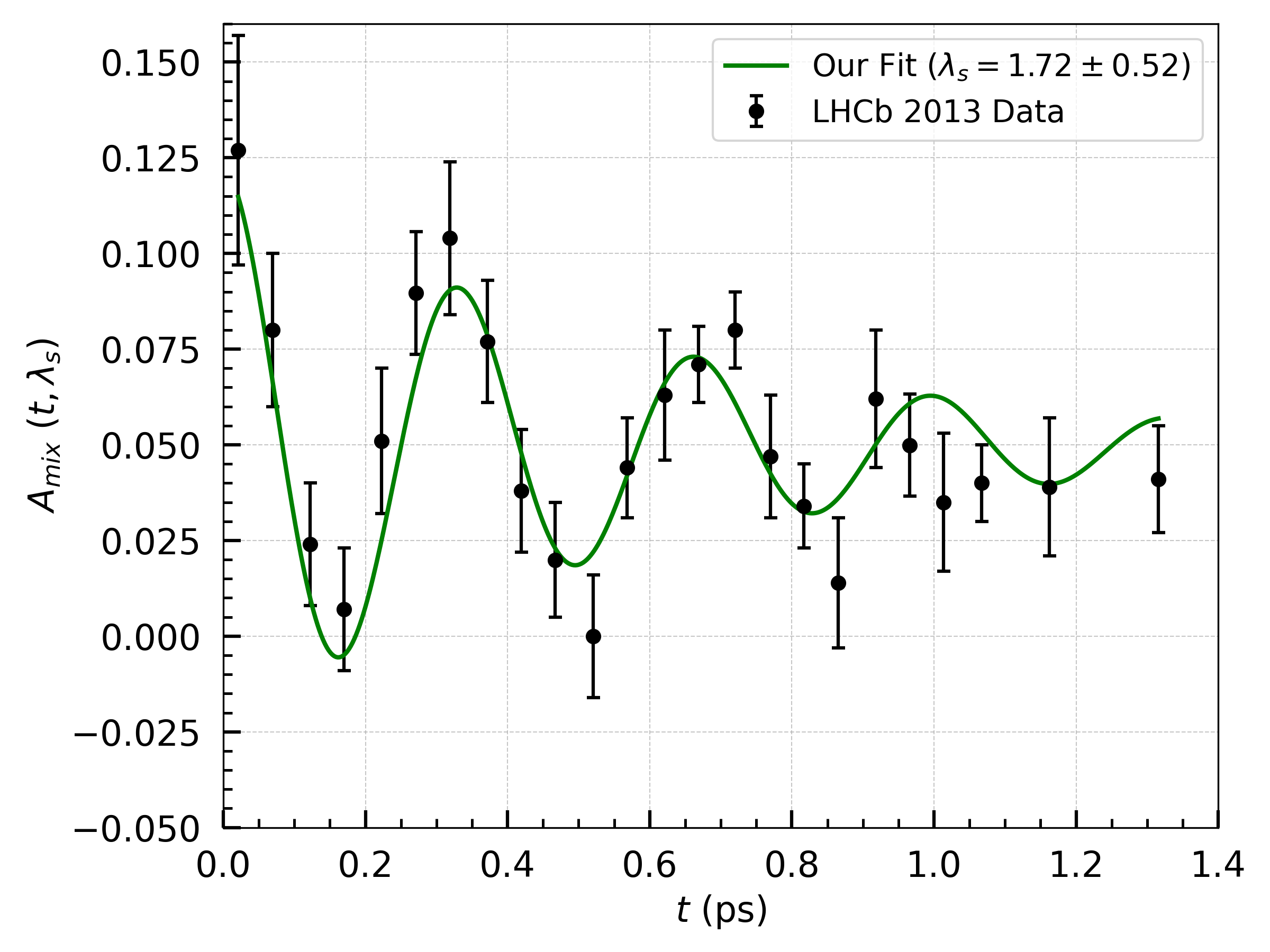}
 \caption{ Time-dependent mixing asymmetry \( A_{\text{mix}} (t,\lambda_s) \) as a function of decay time \( t \) for the \( B_s \)-meson system. The black points represent the LHCb 2013 data, while the green curve shows our fit to the data, including the decoherence parameter \( \lambda_s \). The error bars indicate the uncertainties in the experimental measurements. The value of {$\Delta m_s$} obtained from our fit is $18.65 \pm 0.32 $ \(\mathrm{ps}^{-1}\) when decoherence is not considered, and is $18.85 \pm 0.33$ \(\mathrm{ps}^{-1}\) when decoherence is taken into account.}
\label{fig:delms}
\end{figure}

The mixing asymmetry described in  Eq.~\eqref{amix} depends on the width difference $\Delta \Gamma$ also. However, unlike the \( B_d \)-meson system where \( \Delta \Gamma_d \) is negligible, a nonzero value of \( \Delta \Gamma_s \) has to be taken into account. In the LHCb measurement, we analyzed, \( \Delta \Gamma_{s} \) was treated as nominally constant at its current world average value, with small variations within the known uncertainties having only a marginal impact on the results. We also adopted the same approach for \( \Delta \Gamma_{s} \) in our fits. Given the statistical fluctuations and poor oscillation patterns at larger times, a subset of data points is extracted up to the decay time, where reasonable accuracy can be achieved using our data extraction tools. Despite this constraint, the dataset is adequate for our analysis. To address the precision limitations of the data extraction method, a 5\% variation in the statistical uncertainties is also included in our analysis and the extracted data is shown in Figure~\ref{fig:delms}.

We employ the same analysis strategy used for the $B_d$ system to validate the fitting procedure. Initially, the fits are carried out under the assumption of no decoherence effects. The resulting value of \( \Delta m_s = 18.65 \pm 0.32 \, \mathrm{ps}^{-1} \) with $\chi^{2}_{bf}/ dof$ = 1.74 is consistent at $2\sigma$ level with the published result \( \Delta m_s = 17.93 \pm 0.22 \, (stat.) \pm 0.15 \, (syst.) \, \mathrm{ps}^{-1} \), considering the uncertainty associated with data extraction. Subsequently, the fits are repeated while incorporating the decoherence effect. This fit yields a mixing frequency of $\Delta m_{s} = 18.85 \pm 0.33$ \(\mathrm{ps}^{-1}\) and the decoherence parameter of $\lambda_{s} = 1.72 \pm 0.52$, with an improved $\chi^{2}_{bf}/ dof$ of $1.02$.

Our findings indicate a non-zero $\lambda_s$ at more than \( 3\,\sigma \) confidence level and that its magnitude is much higher than that of $\lambda_d$. 
Hence, an analysis aimed at estimating $\lambda_s$ using the latest data from LHCb has the potential to yield  more compelling results and could confirm a non-zero $\lambda_s$ with greater precision.  The LHCb collaboration is well-positioned to analyze their data directly and provide a precise determination of \( \lambda_s \).


\textit{Conclusion.}---The presence of decoherence can influence several important measurements in the \( B \)-meson system. In this work, we performed a combined fit to the available data for mixing asymmetry and \(CP\) asymmetry in the \( B_d \)-meson system to estimate the decoherence parameter $\lambda_d$. We find that it is non-zero at a \( 6\, \sigma \) confidence level and leads to $4 \,\sigma$ deviations in the central values of $\Delta m_d$ and \( A_{\rm CP}^{\rm mix} \). We also provide the first experimental constraints on the decoherence parameter in the \( B_s \)-meson system from the mixing asymmetry data, where it is also found to be non-zero at the \( 3\,\sigma \) confidence level. The current framework has the potential to strengthen the existing constraints on the value of the decoherence parameter, with further improvements in precision measurements expected at Belle II, additional Run-3 data from LHCb, and forthcoming HL-LHC.\\

\textit{Acknowledgements.}---This work is dedicated to the memory of Prof. Ashutosh Kumar Alok, who tragically passed away during the preparation of this manuscript. His presence is deeply missed and will always be remembered. The work of JK is supported by SERB-India Grant EEQ/2023/000959 and SUS thanks to the Institute of Eminence funding to I.I.T. Bombay from the Government of India. 

\end{document}